\documentclass[11pt]{article}
\usepackage{amssymb,amsbsy,amsfonts,amssymb,amsmath,setspace}
\newcommand{\be}{\begin{equation}}
\newcommand{\ee}{\end{equation}}
\newcommand{\bel}[1]{\begin{equation}\label{#1}}
\newcommand{\bea}{\begin{eqnarray}}
\newcommand{\eea}{\end{eqnarray}}
\newcommand{\beal}[1]{\begin{eqnarray}\label{#1}}
\newcommand{\nn}{\nonumber}

\newcommand{\xp}{x_{\bot}}

\def\d{\partial}
\def\ra{\rightarrow}
\def\Fd{\hbox to 18pt { {\large $\cal F$} \hss {\it~d} } }

\begin{document}

\begin{center}
{\bf\Large\bf
 Parton models and frame independence of
high-energy cross-sections.  }

{\bf O.V.~Kancheli} {\normalsize\footnote{~Email:
 ~kancheli@itep.ru}}
\\[3mm]
{\em Institute for Theoretical and Experimental Physics, \\117218,
Moscow, Russia}
 \vspace{10mm}

\end{center}

 \vspace{-8mm}

  \begin{abstract}
We describe some ambiguities which  take place when on calculates
the cross-sections  in parton models at high energies and the
connected limitations on the asymptotic of high energy amplitudes
that follows from the conditions of boost-invariance of
cross-sections.

It turns out that the resulting constraints are of the same type
as the following from the t-channel unitarity conditions.~  So
that on can suppose that this similarity, by their nature, has
much more general grounds.
 \end{abstract}

\vspace*{5mm}
\setcounter{tocdepth}{2}


\section{\bf  Introduction   }

There are two main theoretical approaches to a study of the
behavior of high energy amplitudes and cross-sections.

In one approach, we directly calculate the amplitudes by summing
the contributions of the Feynman diagrams of the corresponding
field theory or use some effective theory like reggeon diagrams or
various string-like dual models.

In the other - parton like approach to high-energy collisions
  \footnote{~A few useful reviews  ( \cite{grpar}- \cite{kole})~  }
 ~we usually consider separately three main stages of the system
evolution in the process of particles collision. ~~Firstly, one
constructs the quantum states ~$\Psi (P)$ of high energy particle
with momentum $P \gg m$ in terms of superposition
 \bel{pdec}
\Psi (P) ~=~ \sum_n \int_{\{k_i\}} ~f_n(P,\{k_i\}) ~|n,\{k_i\}>
 \ee
 of the n-particle states $|n,\{k_i\}>$
of some ``primary'' constituents - partons with 3-momenta
$\{k_i\}$.~ The ``choice'' of these partons is not unique, and
partons can be bare point like particles, particles with varying
virtuality, QCD color dipoles,  fast string configurations,
distributions of the Coulomb-like fields, etc.~

 The state $\Psi (\vec{P})$ must fulfill the Schroedinger equation~~
$$
\hat{H} ~\Psi (\vec{P}) = \sqrt{\vec{P}^2 + m^2} ~~\Psi
(\vec{P})~,
$$
where the Hamiltonian  $\hat{H}$ is the function of parton
fields,~ so that  $\Psi (\vec{P})$ is the eigenfunction of
$\hat{H}$ with eigenvalues defining  the particles physical mass~
$m$.

~After that one can use such a state $\Psi (P_1)$ to calculate its
interaction  with some low energy target or with other fast
particle  in the state $\Psi (P_2)$ in terms of ``simple''
amplitudes of parton interaction.

There is also the third  stage corresponding to an evolution in
the final state when moving away partons transform and combine
into  physical particles (hadronization...). But often this stage
is not very restrictive, especially when we calculate various
integrated cross-sections. And we will not consider it in this
article.

It is essential that with the energy growing  in most parton
descriptions the structure of the parton state becomes more and
more complicated for all the theories containing vector (like QCD)
and tensor fields (gravity) and  mean parton number in states
$\Psi (P)$and the average transverse size of the region they
occupy grow with~$P$.

When we consider the collision of two fast particles in the parton
states $\Psi (P_1)$ and  $\Psi (P_2)$  at some large $s = (P_1 +
P_2)^2 \gg m^2$ we can choose for this any longitudinal Lorentz
system. But the resulting values of cross-sections of various
processes must not depend from this choice of frame. And this is
nontrivial condition in parton approach, because in different
longitudinal systems (that is for various $P_1$ and $P_2$ at the
same value of $s$) the different parton configurations firstly
meet one another at the moment of particles collision. And,
moreover, by choosing a different system we also can move the
dynamics, from stage one to two and vice versa.

If we make all calculation precisely - with hermitian  Hamiltonian
we probably can be sure that all restrictions coming from
Lorentz-invariance and  the unitarity conditions will be
satisfied. But if we make some approximations, especially dictated
by  phenomenological or pictorial arguments, the unitarity
conditions itself can probably be the only general way to check
that the results are not contradictory.

Various restrictions from the t-channel unitarity are very
essential for the amplitudes describing high energy hadron
interactions, and they are directly taken into account in reggeon
amplitudes  \cite{rft}. But in parton approaches it is not evident
how to take them into account.

In the reggeon field theory and in the dual (string) models the
t-unitarity conditions are automatically fulfilled. But at high
reggeon (pomeron) density such un approach can become unreliable.
The parton approach has no problems with high parton density, but
here there is no direct way how to control possible restrictions
coming from the t-unitarity.

One can hope that the longitudinal Lorentz (boost) invariance of
all cross-sections calculated in a parton approach is in some
sense equivalent to the mean form of the t-unitarity for
multiparticle amplitudes.~
  So, if we calculate any cross-section using the partonic wave
functions $\Psi (P_a)$ and $\Psi (P_{b})$ of fast colliding
hadrons with momenta $P_{a},$ $P_{b}$ \ then  we expect that  this
cross-section must be the same in all longitudinal Lorentz frames
- that is if we calculate the cross-sections using \ $\Psi
(L(\vartheta )P_{a})$ and $\Psi (L^{-1}(\vartheta )P_{b})$, where
$L(\vartheta )$ is a longitudinal boost. It is essential, that in
a parton picture such boosts $L(\vartheta )$ act on hadrons Fock
state very nontrivial changing the number of  partons, etc.

No precise arguments for such general propositions (the boost
invariance for parton cross-sections $\simeq$ t-unitarity) are
known.~ Although it is by itself natural  that the calculations of
cross-sections in the parton picture must give a frame independent
answer.~ Also this is, in particular, confirmed in if we give the
partonic interpretation to reggeon diagrams, by t-cutting them at
various intermediate rapidities, as if we calculate various
multiparticle inclusive cross-sections.

In this article
 \footnote{The material  of this paper partially intercepts with
the article of the author \cite{kan}.}
 we consider some  examples illustrating  how
the requirement of boost-invariance essentially restricts the
structure of high energy collision dynamics. We see that it
restricts in the same way  as it follows from the conditions of
t-unitarity.

 \vspace{8mm}
 \section{\bf   Restrictions on a parton states  from
 the boost  invariance  of high-energy  collision
 cross-sections.\\
~~~Simple Examples}

In this section we illustrate how the requirement of the frame
independence (boost-invariance - BI) restricts the behavior of
high-energy cross-sections calculated in the parton approach.

 We suppose that partons are point like particles with perturbative
interaction and consider here some examples which show how BI
condition works. Also we choose very high energy interactions,
where the mean number of parton in HE state is large, so one can
consider firstly only states with mean number of partons and only
after that take into account corrections from other components of
the Fock wave function of a fast particle. ~So, the picture of
interaction is almost quasiclassical.

We consider the behavior at a boost-transformation of the
inelastic cross-sections $\sigma_{in}$ or of the connected
quantity - the transparency $T = 1- \sigma_{in} = |S|^2$, which is
often more sensitive to the breaking of BI.~
 We choose some frame  where the colliding particles have rapidities
$y_1=y$ and $y_2=Y-y$, where $Y=\ln(s/m^2)$,~ and require that
calculated cross-sections do not depend on~$y$~

 We begin from the simplest parton models of a fast hadron - the rare
parton gas state and of the black disk state.

 \vspace{6mm}
 \subsection{ Collision of a rare gas like parton states}
 \vspace{3mm}

Let us consider the collision of two particles which can be
represented as the partonic clouds that are in a state of a very
rare gas. This is the case usually described by  reggeon diagrams,
that, by their construction, include t-unitarity requirements.
 Let the mean number of partons in colliding hadrons be $n(y)$,
$n(Y-y)$~ and the mean transverse radii of regions occupied by
these partons are~ $R(y)$, $R(Y-y)$, respectively. Then the total
inelastic cross-section can be expressed~as:
 \beal{st1}
  \sigma_{in} (Y) ~=~ \sigma_0~ n(y)~n(Y-y)~~-~~~~~~~~
  ~~~~~~~~~~~~~~~~~~~~~~~~~\nn \\
  ~~~-~c_1 ~\sigma_0~ n(y) n(Y-y)~\Big(~ \frac{\sigma_0 ~n(y)}{R^2(y)}
  ~+~ \frac{\sigma_0 ~n(Y-y)}{R^2(Y-y)} ~+~ \\
  ~+~ \frac{\sigma_0~ n(y)~n(Y-y)}{R^2(y)+ R^2(Y-y) }
  ~\Big) ~+...~~,~~~~~~~~~~~~~~~~~~ \nn
 \eea
where $\sigma_0$  is the parton-parton cross-section, ~$c_1 \sim
1$.~ The first term in (\ref{st1}) corresponds to a collision of
at least one pair of partons. The next terms describe corrections
from screening and multiple collisions~
 \footnote{The cross-sections of local interactions of
point-like particles decrease as a function of their relative
energy as $\sigma_0(s) \sim 1/s$. As a result in  (\ref{st1})
enter, in fact, only the numbers of low energy partons $n(y) ,
n(Y-y)$ of the colliding particles in this coordinate system. } .

For the rare parton gas one can at first approximation neglect
multiple collisions and screening, that is to leave only the first
term in~(\ref{st1}). Then, from the requirement of the
independence of ~~$\sigma_0~ n(y) n(Y-y)$~~ on ~$y$~ follows the
unique solution for ~
  \bel{n0}
 n(y) =~n_{0} e^{y \Delta_0}
  \ee
with some real constants $n_{0}$, $\Delta_0$.~ The following from
(\ref{n0}) behavior of~~
  \bel{n1}
  \sigma_{in} (Y) =  \sigma_0~n_0^2~e^{Y \Delta_0}
 \ee
in the elastic amplitude corresponds to a regge pole in the
complex angular momentum plane (and not to a cut or some more
complicated regge singularity~). And this condition follows \cite
{GrPo} in a relativistic Regge approach only from the 2-particle
t-unitarity of the elastic amplitude.

Note that the coefficient in (\ref{n1}) is in fact factorized -
for the collision of different particles a+b one must $n^2
\rightarrow n_a n_b$~. ~This factorization in regge approach also
follows from t-unitarity.

Moreover, it is interesting to consider \cite{Kaid} the behavior
cross-section $\sigma_{in}$  with the definite impact parameter
$B$, normalized so, that
 \bel{st1b}
 \sigma_{in}(Y)~=~ \int~d^2 B~  \sigma_{in} (Y,y,B)~~.
 \ee
In this case  the analog of the first term in (\ref{st1}) can be
represented as
 \bel{ast2b2}
 \sigma_{in} (Y,y,B)~=~ \sigma_0 \int d^2 x_{\bot}
          ~\rho(y,|x_{\bot}|)~\rho(Y-y,|B-x_{\bot}|)~~~~,~~~~
 \ee
where $\rho(y,x_{\bot})$ is the transverse parton density~ $(~n(y)
= \int d^2 x_{\bot}\rho(y,x_{\bot}) ~)$ . ~Then from the frame
independence of the $\sigma_{in} (Y,y,B)$ the form of transverse
parton density $n(y,x_{\perp})$ can be essentially restricted.
 The condition of y-independence can be writhen as
  \bel{nn0}
 \frac{\d}{\d y} ~\sigma_{in} (Y,y,B)   ~=~ 0~.
 \ee
Going here to conjugate to $\xp$ variable
 $$
 \rho(y,\xp)  = \int d^{2} k
 \cdot  e^{ik \xp} ~\tilde{\rho}(y,k)
 $$
we come from (\ref{nn0}) to the equation
$$
  \frac{\d}{\d y}~ \big(~ \tilde{\rho}(y,k)
                 ~\tilde{\rho}(Y-y,k)~\big) ~=~ 0~,
$$
which has the solution
$$
\tilde{\rho}(y,k) = f_{1}(k) \cdot  e^{ yf_{2}(k) }
$$
and then as a  result
 \bel{nn1}
   \rho(y,\xp) \sim \int d^{2} k ~ e^{ik \xp}~
  f_{1}(k)   ~e^{ yf_{2}(k) }~,
 \ee
where $f_1$, $f_2$ are arbitrary functions of $k$. For $y \ra
\infty $ the integral in (\ref{nn1}) can be taken by the steepest
decent method, so that only the neighborhoods of zeros of $\d
f_2(k)/\d k$ are essential. Then from the positivity of the parton
density $\rho$ it follows that $f_2$ is positive and so the
dominant contribution must come from the region $k \sim 0$,
otherwise $\rho(y,\xp)$ will oscillate in $\xp$. So in the
essential region $f_2(k) \simeq c_1 -c_2 k^2, ~~c_2 > 0$, and
estimating the integral  (\ref{nn1}) we come to the expression for
the density of low energy partons
  \bel{nrp}
\rho(y,\xp) ~\sim~  y^{-1}~
    e^{\big(c_1 y -\xp^2/4 c_2 r_0^2 y \big)} ~~,~~~c_2 > 0~.
  \ee
The expression (\ref{nrp}) corresponds to the Gauss form of parton
distribution in  $\xp$ which usually results from the diffusion of
partons in $\xp$ plane during the parton cascading. The mean
radius of a low energy parton cloud  $R(y)~\sim r_0 \sqrt{~y}$ ~is
also fixed here only from the condition of the frame
independence.~~
 In the elastic  amplitude the Eq. (\ref{nrp}) corresponds to the
contribution of the regge pole with the trajectory $\alpha(t) =
1+\Delta + \alpha' t$~, ~where $\Delta = c_1 ,~\alpha' = c_2$.

If we make the next step and impose the condition of $y$
independence on the sum of two terms in the right side of
(\ref{st1}) and assume that the correction to (\ref{n0}) is small,
we become instead of (\ref{n0}) the corrected expression
  \bel{ny2}
  n(y)=n_{0} ~e^{\Delta_0 y}  -~
    a_2 ~n_0^2 \frac{\sigma_0}{R^2 (y)} ~e^{2 \Delta_0 y}
    ~+...~~~~
 \ee
From here it is simple to conclude that
 \bel{n2}
  \sigma_{in} (Y) =  n_0^2~ \sigma_0 e^{(\Delta_0 Y)} ~-~
 ~n_0^4 \frac{a_2 \sigma_0^2}{R^2 (Y)} ~e^{2(\Delta_0 Y)}~~.
 \ee
The second term in (\ref{n2}) corresponds to the the contribution
of two reggeon cuts, whose structure is almost complectly fixed
here from the boost-invariance. The arbitrary coefficient $a_2 > 1
$, depends on the weight of the diffractive amplitudes entering in
the two regeon emission vertex. The possible next terms in
(\ref{ny2}), corresponding to higher regge cuts,  can be found in
the same way by iterative applying the boost-invariance condition
to the  combinations of screening terms in the expression
(\ref{st1}) for~$\sigma_{in} $.

Thus, it can be seen that for rare parton states we come to the
restrictions on there structure that arise from the reggeon
diagrams and are defined by the t-unitarity.
   \vspace{3mm}

At the end of this section note that at at all currently available
energies the dominant high energy hadron interactions are well
described by the regge approach with the soft pomeron exchange and
the respective cuts. This directly corresponds to the Gauss-like
parton distribution consistent with the parton frame independence.

 \vspace{6mm}
 \subsection{Collision of a black  disks}
 \vspace{3mm}

Now let us consider the opposite limiting case of colliding parton
clouds, when the mean parton density is very high and partons fill
a transverse disk with the radius  $R(y)$ depending on particles
energy $E = m e^y$. ~Then the total inelastic cross-section can be
determined from purely geometrical conditions - it is defined by
the area of an impact parameter space, corresponding to the
overlapping of the colliding black disks :
 \bel{bd1}
 \sigma_{in} (Y) ~=~  \pi~ \Big( R(y)+R(Y-y) \Big)^2~.
 \ee
From the condition  of independence of the right  side of
Eq.(\ref{bd1}) on $y$~  evidently follows the unique solution for
 \bel{fr1}
  R(y) ~=~ r_0 \cdot y + r_1
 \ee
It is interesting that in this case we immediately come directly
to asymptotically constant cross-sections (when $r_1 = 0$),~ or to
the Froissart type behavior of cross-sections~
  \bel{froibd}
 \sigma_{in} (Y) \simeq \pi r_0^2 Y^2  + \pi r_0 r_1 Y +\pi
 r_1^2~~.
  \ee
Here, in the Froissart case the elastic cross-section is
diffractive and $\sigma_{el} = \sigma_{in}$.
 ~Also the terms $ ~\pi r_0 r_1 Y $~ in (\ref{froibd}) correspond
to a diffraction generation as is natural in the Froissart case.

 \vspace{6mm}
 \subsection{Collision of the parton grey disks}
 \vspace{3mm}

The real parton disk ~(even at $Y \gg 1$)~ cannot be absolutely
black because the parton density at every particles energy is
finite. Besides that the local parton density fluctuations  also
can lower the parton density in the individual events and this
leads to the grow of the locale transparency of such disks.~ For
such parton disks the conditions of BI can lead to rather strong
restrictions on the structure of ``grey'' parton states and their
interactions.

Firstly, consider the collisions of grey disks with some constant
grayness - when the mean transverse parton density is stabilized
at some fixed value and do not grow with energy i.e. the local
disk transparency also does not change with energy
 \footnote{One can expect this type of the behavior in the (2+1)D QCD,
which is soft and if here the parton saturation takes place
\cite{kan}.}.
 Then it is easy to see that the condition of the
boost invariance can at all not be fulfilled for such  models.

In the lab.frame of one particle the transparency
$$
T_{lab}(Y,B)=const(Y) ~~~~~at~~~~~ Y \rightarrow \infty ~~,
$$
because at all $B$ only a finite number ($\sim 1$) of partons must
penetrate through the grey parton disk of the other fast
particle.~

And in the center of the mass system at the same impact parameter
the large number of partons ~$N_{12}$~  must penetrate. For the
grey disk $N_{12} \sim S_{12}(y,Y,B)$ - the transverse area of two
disks overlapping region.~~ And for growing with $Y$ disk radius
the~ $S_{12}$ also grows. For the Froisart type growth we have
$S_{12} \sim Y^2$ at $B \ll R(Y)$. ~ Then in systems close to the
center of mass
 $$
 T_{scm}(Y,B) \sim
 e^{(-c N_{12})} \sim  \exp{(-c Y^2)}
    \rightarrow 0 ~~~~,~~~~~~~~c\sim
 1~~~.
 $$
Therefore, the case of a grey disk with a constant (or a slowly
($\sim 1$) varying~) parton density should be probably excluded.

 \vspace{1mm}
In all more or less realistic situations, such, for example that
one can expect in the QCD, the parton disk can have a grey parton
border, even when the inner parts of the disk become almost black.
In this case the average parton density  can be roughly
represented as
 \bel{ro3}
   \rho(y,\xp) ~\simeq~
  \rho_d (\xp)  ~\theta (R(y) - \xp )  ~+~
  \rho_0 ~\theta ( \xp  - R(y)) f(\xp) ~~,
 \ee
where $\rho_d (\xp)$ describes the behavior of the parton density
in the inner part of the disk, and where the grey border has the
width~~$\lambda (y) \ll R(y)$. ~In this border the parton density
varies from the high value (almost black) to small one. For
example, it can have the form
 \bel{bordis}
 f(\xp) ~\simeq~ e^{-( \xp -R(y))/\lambda (y)}~~.
 \ee
For collisions with an impact parameter $B < R(Y)$, when colliding
disks stuck with their almost black parts, we possibly can have a
boost-invariant behavior of~ $\sigma_{in}$. ~But for collisions
with $B-R(Y) \sim \lambda (Y))$, when the discs collide with their
grey edges, the situation is different.

 In the Lab frame of one of particle the transparency
$T_{lab}=const \sim 1$, because here only some ($\sim 1$) partons
must penetrate without interaction through the grey edge of the
large disc.

And in the arbitrary system at the same impact parameter the
transparency is
$$
   T(y,Y-y,B) \sim e^{-N_{12}(y,Y-y,B)}
   \sim  e^{- S_{12}(y,Y-y,B)/r_0^2}~~,
$$
where  $N_{12}(y,Y-y,B)$ is the average number of parton
interactions during the collision and $S_{12}(y,Y-y,B)$ is the
area of the two disks intersection region. This region has a form
of elongated figure whose width is  $\sim \lambda(y)$ and the
length $l(y) \sim \sqrt{R(y) \lambda(y) }$ for $y \lesssim Y-y$.~~
So, for such B the two disks intersection  area is
 $$
   S_{12}(y, Y-y, B \simeq
    R(Y)+\lambda) ~~\sim~~ R(y)^{1/2}*\lambda(y)^{3/2}~~.
 $$
In the center of mass system this gives for $Y \gg 1$
 $$
  T_{scm} \sim \exp ( - c (R(y)/r_0)^{1/2} )  \rightarrow 0
 $$
even for parton disks with $\lambda \sim const(Y)$, although the
width of the grey border can also grows together with the disk
size~
 \footnote{One can expect \cite{kan} that for a realistic parton
disc due to border shape fluctuations the mean width of the grey
zone grows with Y as $\lambda \sim \sqrt{Y}$}.
 ~Therefore, for the border collisions with such  $B$ and $Y$ we
have no boost-invariance of T ,  and this conclusion in fact
almost does not depend on the explicit form of the border
distribution $f(\xp)$. Probably, the only exception is the Gauss
type distribution of the parton density when the whole disk has
the ``structure of border''.

\vspace{6mm}
\subsection{ Particle to heavy nuclei interaction }
 \vspace{3mm}

A slightly different type of restriction on the parton structure
follows from the boost invariance if we consider the high energy
collision of a particle p (for example a proton, a pion or any
test color dipole) with heavy nucleus $(A \gg 1)$.

To see  this we compare the estimate of transparency T in Lab
frame of nuclei and the Lab frame of p. Also we  choose Y not very
large, but so that $Y \gg \ln A$, and consider a collision at $B =
0$. In fact, in this case we have a collision of ~p with a long
($\sim A^{1/3}$) tube of nucleons, and we want to calculate the
probability of the passage of p through A without interaction.

First, consider the p $\bigotimes$ A collision in the Lab frame of
p. For such an Y due to the Lorentz contraction of the moving
nuclei all soft partons of the fast nuclei  are placed in a tiny
transverse region of  the longitudinal size $\sim 1/m$. And if the
parton saturation take place, the number of soft partons $N_A$
interacting with p should almost not depend on A, because all
``additional'' soft partons coming from different nucleons in the
A-tube are absorbed one by another. Therefore, one can expect that
the transparency in the p-lab. frame is
   \bel{tp1}
   T_p ~\sim~ e^{ - N_p(Y)}~~.
  \ee
On the other hand, to calculate T in the Lab frame of A at the
same B and Y  we must find the probability that fast particle p
penetrate without interaction trough the $ A^{1/3}$ long tube of
nucleons. Here one can expect that
  \bel{tA1}
  T_A ~\sim~  e^{- c(Y) A^{1/3} }
  \ee
Because in such a ``thought experiment'' we can arbitrary choose Y
and A and the distance between the nucleons in the tube, we come
to an apparent contradiction with the frame independence. This
means that some constrains must be imposed on the parton dynamics.
The simplest way is to suppose that there is almost no parton
saturation in the A-tube. Or on the contrary - that some kind of
the mechanism works, which makes the interaction of a fast p
particle with the nucleus otherwise dependent on A.

Possibly some indications on the causes of this inconsistency can
be found if we consider the regge description of this reaction,
where we can calculate $\sigma_{in}(Y,B) = 1 - T$ for a large A
and not to a large Y.

If we take for a single pomeron exchange in the p~$\bigotimes$~A
reaction the amplitude
$$
v(y,b) \sim i g^2  A^{1/3} \exp{(\Delta y  -b^2 /4\alpha' y)}
$$
and consider firstly the simple eiconal case which corresponds to
a situation without parton saturation we become for the
corresponding S-matrix $ S(Y,B)  = \exp{ (i v(Y,B))}$, and this
gives for the transparency
  \bel{tA2}
     T(Y,B=0)  ~=~ |~S(Y,B=0~|^2 ~\sim~
     \exp{ \Big( - 2 g^2  A^{1/3} e^{\Delta Y}   \Big)} ~~.
 \ee
The simplest way to take into account something similar to the
parton saturation is to include into the single pomeron exchange
amplitude the pomeron cascading from the side of A-vertices. ~So
that from the p-side the pomeron line joins to p and from the A
side the pomeron line branching  joins to many nucleons. This
corresponds to the new amplitude
   \bel{vvam}
 v ~\rightarrow~    \tilde{v} ~=~
 \frac{v}{1 - i \frac{r}{g \Delta} v }~~,
  \ee
were $r$ is the 3-pomeron vertex.~ In this case for large
$A^{1/3}$ and $B~=~0$ the amplitude $ \tilde{v}$ is stabilized at
the value $| \tilde{v}|  = g\Delta/r$. And, therefore, the
corresponding transparency approaches to
 \bel{tp2}
  T(Y,B=0)  ~=~  \exp{\big( -2 | \tilde{v}| ~\big) } ~=~
             \exp{\Big( - 2 g\Delta/r \Big)}~~.
  \ee
Comparing the expressions ( \ref{tp1} ) with (\ref{tp2}) and
(\ref{tA1}) with (\ref{tA2}) we see their similarity, but this
unfortunately does not help to find the right answer, because the
simple expression like (\ref{vvam}) dos not take into account
various pomeron interactions in the $\tilde{v}$-cascade and also
the other pomeron interactions in eiconal multipomeron diagrams
 \footnote{ Note that approximately the same inconsistency appears
if we consider the heavy  A~$\bigotimes$~A  interaction end
compare the estimates of T in the Lab frame and in the CM system
}.

\vspace{6mm}
\subsection{Possible boost-invariant parton density
distributions in a grey disk }
 \vspace{3mm}

In fact, in the case of asymptotically growing cross-section all
parton distributions corresponding to real theories like QCD will,
probably, lead to the grey dick.  And it is interesting to find
the sensible examples of parton distributions that correspond to
the boost-invariant T.

Let us consider collisions of particles with some  parton
distribution $\rho(y,\xp)$ and try to find the minimal conditions
on the form of $\rho(y,\xp)$ for which cross-sections are
boost-invariant

With the exponential precision the  transparency can be expressed
as:
 \bel{trr}
T(Y,y, B) ~\sim~ \exp \Big(~ - N (y,Y-y, B) ~\Big) ~~~,
 \ee
where
 \bel{taurj}
   N (y, Y-y, B) = \sigma_0 \cdot \int d^{2}b
  \cdot \rho (y,|b|) \cdot \rho (Y-y,|B-b|)
 \ee
is proportional to the mean number of the parton scattering when
two \Fd penetrate one through another during their collision at
the impact parameter~B.

Because the expression (\ref{taurj}) has the same structure as
(\ref{ast2b2}) one can repeat here the calculation given above.
Then we find that  the expression ( \ref{taurj} ) can be boost
invariant only for some very special   Gaussian form of parton
density $\rho$ inside the disk :
 \bel{redf}
  \rho (y, \xp ) ~\sim~ \rho_0 ~\frac{1}{y}~e^{\Delta y  -
    \xp^2/y r_0^2}~~,~~
 \ee
This corresponds to the distribution arising in the parton cascade
when partons only split and do not join. The same answer (Eq
(\ref{nrp})) for $\rho (y, \xp )$ was found for the rare parton
systems - but here the density can be arbitrary high. In the
connected elastic amplitude it corresponds to a regge pole
exchange with the intercept $\Delta$.~ In fact, the expression
(\ref{redf}) for $\Delta > 0$ corresponds again to almost black
disk (but without parton saturation~!) of the radius $r_0 \Delta
~y$~ with a thin grey border, because the parton density changes
here very fast from a small to a big values at the distances
$\delta \xp \sim r_0/\sqrt{\Delta}$.

In general case one must take into account that partons in the
colliding disks can have different virtualities  $u \sim \ln
k^2_{\bot}/m^2$~, where the parton density $\rho (y,b,u)$ has now
nontrivial dependence on $u$.  Partons with large $u$ are more
strongly localized in transverse coordinates and their interaction
cross-sections $\sigma(u_1,u_2)$ usually decrease for large $u_i$.
The expression for the transparency in the process of collision of
two parton disks  has again the form (\ref{trr}), where the mean
number of parton interactions during the collision is given by the
following generalization of (\ref{taurj})
 \bel{tauruj}
   N (y, Y-y, B) =   \int d^{2}b \int du_1 du_2~
 \sigma(u_1,u_2) \cdot  \rho (y,|b|,u_1) \rho (Y-y,|B-b|,u_2)~.
 \ee
In this case the restrictions on the form of $\rho (y,b,u)$ coming
from the frame independence condition ~$(\d/\d y)N (y, Y-y, B) =
0$~ are not so strong as for~(\ref{taurj}~-~\ref{redf}).

 If the parton cross-sections that enter (\ref{tauruj}) can be
approximately factorized as
 $$
\sigma(u_1,u_2) \sim \ell(u_1)\cdot \ell(u_2) ~~,
 $$
then  the condition for the boost invariance of $N$ can be reduced
to the more simple  equation
 \bel{ures}
    \int du~ \ell(u) \rho (y,b,u) ~=
      \rho_0 \frac{1}{y}~e^{\Delta y  - b^2/y r_0^2}
 \ee
In this case the  form of $\rho (y,b,u)$  for some interesting
models are again almost completely fixed.~ For example, so is the
superposition of grey saturated disks with different virtualities
$$
\rho (y,b,u) \sim  \varphi(y,u) ~\theta (r_1 \chi(y) - b u^a )~,
$$
so that the mean radii of these disks $r_1 \chi(y)/u^a $ decrease
 \footnote{In QCD the radii of hard subdisks can grow as $\sim
y/\sqrt{u}$, and this corresponds to $a=2$}
 with growth of~u. Here, from equation(\ref{ures}), one can find the
explicit expression for
  \bel{phy}
  \varphi(y,u) ~=~ \varphi_0 ~\frac{  e^{\Delta y} }
                        {\ell(u)~u^{2a}~y}
       ~\exp{\Big( - c_2 \frac{\chi^2(y)}{y~u^{2a}} ~~,         \Big)}
 \ee
where $\varphi_0,~ a,~ \Delta,~ c_2$ and functions $\ell(u),~
\chi(y)$ can be chosen arbitrary.
 If we choose $\chi (y) =~\chi_0 y $ ~so to have the Froissart type
of the growth of disk radius we will have from (\ref{phy}) for the
disk density
 \bel{phyf}
  \varphi(y,u) ~\sim~ ~\frac{  e^{\Delta y} }
                        {\ell(u) u ^{2a} y}
       ~\exp{\Big( - \tilde{c_2} \frac{y}{ u^{2a}}~~.         \Big)}
 \ee
For large $u$ it is natural to expect that $\ell(u) \sim e^{-c u}$
and therefore the mean density of hard subdisks will grow with u
and y.

\vspace{6mm}
\subsection{Corrections to the mean picture  from
 a big fluctuations in the colliding states }
 \vspace{3mm}

To discuss if the boost-invariance can be somehow restored also
when the mean parton density $\rho (y, \xp )$ is not of the Gauss
form (\ref{redf}) on must take into account all essential parton
configurations,   and also these ones that are very far from the
mean one.  In this case, one can hope that in different frames the
main contribution into cross-sections comes from some different
parton components so to compensate the variation of the
contribution of the mean states. Here especially interesting can
be the rare components of $\Psi (P)$.
 In the Fock state of a fast particle such rare parton
configurations contain a relatively small number of partons and
therefore it can give large contribution to the transparency and
compensate the boost non-invariance of $T$ and other quantities in
the mean density states.~ Such configurations can mainly arise due
to large fluctuations in the initial stages of the patron cascade.
CM one can ask for such a parton component ~$|~bare >$ for a fast
hadron that does not contain a black disk at all and interacts
slowly (or does not interact at all).~
 We can schematically represent such a state of fast particle~:
$$
\Psi (P \ra \infty) ~\simeq~ f_d~|disk > +~
        f_b~|~bare>~,~~~ f_d \gg f_b ~,
$$
where $f_b$ is the amplitude of the rare component $|bare >$ ~and
$| disk> $ is the superposition of ``big'' parton components that
gives the main contributions in a various cross-sections.

The probability for a fast hadron to be in the rare state is
~$w(y)\sim | f_b|^2$. In this case the expression for the
transparency can be generalized to~:
 \beal{trimp}
T(y,Y-y) ~~\simeq~~ T_{mean}(y,Y-y)
 +~ \tau_{b d} \cdot \big( w(y) + w(Y-y) \big) ~~+ \nn \\
 +~\tau_{b b } \cdot w(Y-y)) \cdot w(y)~,~~~~~~~~~~~
 \eea
where the transparencies of the rare component $\tau_{b d}$ and
$\tau_{b b }$  can be finite and do not decrease with growth of
$y$.

The term $T_{mean}(y,Y-y)$ coming from the $|disk > \bigotimes
|disk >$ interaction is not boost-invariant and it can be $const(Y
\rightarrow \infty$) in the lab.frame for a saturated (grey) disk,
and very small in csm.

The two last terms in (\ref{trimp}) coming from the ~$|bare> \cdot
~|disk>$~ and ~$|bare> \cdot ~|bare>$~ components can dominate and
so can make $T$ boost invariant. But it is possible only if $w(y)$
is approximately constant for high~$y$. Various estimates of
$w(y)$ lead to a decreasing function of the type
$$
 w(y)\sim \exp  (-\gamma \cdot y)
$$
for the case of the growing total cross-section.~ It corresponds
to the choice at every rapidity stage of such an evolution
direction, that does not increase the parton number~
 \footnote{Such a behavior of $w(y)$ can also  be found from the
boost-invariance condition applied to the behavior of some hard
cross-sections. See Eq.(\ref{dsdtp} - \ref{dsdtp1}) } .
 Such a behavior of $w$  leads to the expression
 \bel{trgray}
  T(y,Y-y) ~\sim~  \tau_{b d} \cdot \big(~
      e^{-\gamma~ (Y- y)} ~+~ e^{-\gamma~ y} ~\big)
      +  \tau_{b b } \cdot e^{-\gamma~ Y} ~,
 \ee
 corresponding to the collision of the rare state  ~$|~ bare
>$ with other particle.  ~Such contribution to $T$ is $y$
dependent, and therefore on this way the frame indeprndence  can
also not be restored.

 \vspace{6mm}
\subsection{Collision of parton disks in the case of
particles moving in the same direction }
 \vspace{3mm}

When we impose the condition of the independence of the
cross-sections $\sigma_{in}(Y,y,b)$ on the choice of system (i.e.
on $y$), we can choose the values of $y$ not only in the interval
$0<y<Y$, ~i.e. between Lab and center of mass (~CM~) systems. ~But
let us also consider systems with $y<0$ and $y>Y$ , when both
parton disks move in one direction.

This, in principle, can lead to additional constraints on the
amplitudes. But in this case, at first glance, paradoxes may also
arise when estimating the probability of the interaction.
Especially this is seen in the case of the growing cross-section.

To illustrate this let us consider the case of colliding Froissart
type disks when their radii $R(y)$ and $R(Y-y)$ grow with the
particles rapidity as $R(y) = r_0 y$ and $R(Y-y) = r_0 (Y-y)$, and
estimate the behavior of  the inelastic cross-section with the
definite impact parameter $\sigma_{in} (Y,y,B)$ .

We chose $B > r_0 Y$. In this case, when ~$0 < y < Y$,~ the parton
disks pass one by another without interaction. But this is only if
they move towards each other because here $B > R(y)+R(Y-y)$ and
therefore $\sigma_{in} = 0$. But if we at the same $B$ choose the
system so that disks move in the same direction and so that $y \gg
Y \gg 1$ then disks will overlap when one disk will go through
another. And therefore partons from one disk can interact with
partons from another disk.

But it is essential that in such disk interaction no new particles
can be created. Indeed, if in this case a particles can be
created, their momenta will be small ($\sim m$) in this system.
And the creation of such a particle in CM system would correspond
to a creation of a particle with energy $\sim m e^{y}$, where $y
\gg Y$ and this is forbidden by the energy-momentum conservation ;
so $\sigma_{in} = 0$.

From the other hand, the exchange of particles between these discs
with an approximate momentum conservation ~(or, with the exchange
of small transverse momenta)~ can give a contribution to their
elastic scattering and which  comes here  also from large
transverse distances ~(~$B > R(Y)$~).

The parton wave functions of these `` intersecting disks'' can be
entangled one by another by such a mechanism, and also the
conversion of a pure state to a mixed one for every disk can in
principal take place. ~There is here probably no contradiction
with the parton picture, since there is no way to distinguish
between low energy partons in the wave function (\ref{pdec}) and
the close energy partons from vacuum fluctuations.

The entanglement between states of  disks in such a collisions is
proportional to their area. This  suggests that these discs have
entropy $\sim$ their area ($\sim$ the number of low energy
partons), i.e. $\sim y^2$ in this case of the Froissart type
growth of cross-sections.

 \vspace{6mm}
\subsection{Limitations on the dynamic of a hard
elastic scattering  }
 \vspace{3mm}

In the field theory the high energy hard elastic scattering of
point-like particles leads usually to the power behavior of
elastic cross sections
$$
 d\sigma_1^{el} (s,t \simeq -s/2) /d t  \sim  1/s^a   ~,~~~~~
$$
For the scattering of particles  composed from n constituents with
approximately equal momenta we have
$$
 d\sigma_n^{el} (s,t \sim -s/2) /d t  \sim
       \mu^{-4(n-1)}(d\sigma_1 ( s/n^2,t \sim -s/n^2) /d t )^n~~.
$$
But the mean state can contain the growing number of partons and
the direct application of this expression leads to a small
contribution. In this case the main contribution to $ d\sigma /dt$
can come from the rare parton configurations containing the
minimal number of partons (when both particles are in a ``bare''
state). Then, the cross-section of particles in the system, where
$~s=m^2 e^Y$ and the energies of colliding particles are ~$me^y ,~
m e^{Y-u}$, can be represented as
 \bel{dsdtp}
 d\sigma^{el} (s,t \sim -s/2) /d t  \sim
 \big( ~d\sigma_0 (s,t \sim -s/2) /d t ~\big)^{n_0}  ~w(y) w(Y-y)~~,
 \ee
where $w(y)$ is the probability that particle with energy $me^y$
is in the bare state, and $n_0$ - the number of ``valent''
components in the bare state ($n_0 \simeq 2 \div 3$ for meson
$\div$ baryon).~~
 It follows from the boost-invariance of (\ref{dsdtp}) that
 \bel{dsdtp1}
  w(y) \sim e^{- 2 c y}
 \ee
This condition essentially restricts the behavior of the
asymptotic of hard scattering and, in particular, gives the
information about the amplitude (~$\sim  \sqrt{w(y)}$ ) of the
bare component of $\Psi (P)$.

The similar limitation follows from the consideration of the
asymptotic cross-sections of two particle reactions with exchange
of quantum numbers (such as $\pi^- + p  ~\rightarrow~ \pi^0 +
n$).~ Here again, the dominant parton configuration contributing
to such reactions must contain the minimum number of partons. So
again, we have the factor $w(y) w(Y-y)$ in the cross-section.
Additionally, there is the factor of type  $ e^{- 2 g y}$
connected with the probability that this  parton configuration
contains also the small energy parton with ``needed'' quantum
numbers.~ Therefore, from the frame independence of amplitudes of
such reactions we also come to the condition (\ref{dsdtp1}). And,
if interpreting in terms of the exchange of some nonvacuum reggeon
we come to estimate  their intercept as $\alpha(0) \simeq 1 - c -
g$~.

\section{\bf  Summary }

The main aim of this note was to illustrate that the condition of
boost-invariants essentially restricts the behavior of high energy
cross-sections calculated in parton approaches. And the form of
resulting constrains is of the same type as coming from the
t-channel unitarity condition.
  So that one can suppose that this similarity, by their nature, has
much more general grounds.

Such a condition  works especially  effectively in the case of
growing with energy  cross-section, that is, just  when the
t-unitarity conditions for amplitudes is complicated to apply -
because here the multiparticle exchange becomes important.~
 In this case the resulting restrictions on the  asymptotic
behavior are rather strong and can, in principle, exclude some
popular models.

\end{document}